\documentclass[11pt,a4paper]{article}
\usepackage{graphicx} 
\usepackage[blocks]{authblk}
\usepackage{hyperref}
\usepackage[
dvipdfm,truedimen,%
top=30truemm,bottom=30truemm,%
left=25truemm,right=25truemm]{geometry}
\usepackage{color}

\begin{document}
\pagestyle{empty}

\title{\bf Fission fragment distributions of neutron-rich nuclei based on Langevin calculations: toward r-process simulations}

\author[1]{Mizuki~OKUBAYASHI$^*$}
\author[1,2]{Shoya~TANAKA}
\author[1]{Yoshihiro~ARITOMO}
\author[1]{Shoma~ISHIZAKI}
\author[1]{Shota~AMANO}
\author[3,4]{Nobuya~NISHIMURA$^\dagger$}

\affil[1]{Graduate School of Science and Engineering Research, Kindai University, Higashi-Osaka 577-8502, Japan}
\affil[2]{Advanced Science Research Center, JAEA, Tokai 319-1195, Japan}
\affil[3]{Astrophysical Big Bang Laboratory, CPR, RIKEN, Wako, Saitama 351-0198, Japan}
\affil[4]{Division of Science, NAOJ, Mitaka, Tokyo 181-8588 Japan}

\affil[ ]{$^*$Email: m.okubayashi39@gmail.com}
\affil[ ]{$^\dagger$Email: nobuya.nishimura@riken.jp}

\date{}

\maketitle

\begin{abstract}
The nuclear fission of very neuron-rich nuclei related to the r-process is essential for the termination of nucleosynthesis flows on the nuclear chart and the final abundances. Nevertheless, most of the available fission data for the r-process calculations are based on theory predictions, including phenomenological treatments. In this study, we calculated a series of nuclear fission distribution for neutron-rich nuclei away from the $\beta$-stability line. As most of these nuclei are experimentally unknown, we are based on theoretical calculations based on the dynamical fission model with the Langevin method. We performed fission distribution calculations for neutron-rich actinoid nuclei, applicable to the r-process nucleosynthesis simulations. In the present paper, we compared the obtained mass and charge distributions with experimental data. We also show the results of the systematic behaviour of mass distribution for neutron-rich U and Fm isotopes.
\end{abstract}

\section{Introduction}

Nucleosynthesis by the rapid-neutron-capture process, {\it r-process}, represents for cosmic origin of the heaviest elements (e.g., gold and uranium) beyond the iron-group peak. Although several astrophysical scenarios have been proposed, the mechanism of the r-process is not completely understood (for a recent review, see \cite{2019JPhG...46h3001H, 2020PrPNP.11203766A, RevModPhys.93.015002}). One of the main reasons is large uncertainty in the nuclear-physics properties of very neutron-rich nuclei, e.g., neutron capture rates and several decay half-lives. To determine nucleosynthesis flows on the r-process ``path'', the $\beta$-decay and the neutron capture (strongly depending on theoretical mass prediction) are significant. Nuclear fission is a key ingredient of the termination of and the final abundance distribution if r-processing is strong enough to reach actinoids.

Nuclear fission, therefore, plays an essential role under certain r-process conditions \cite{2013PhRvL.111x2502G, 2015ApJ...808...30E, 2016ApJ...816...79S, 2020ApJ...896...28V}, in particular, in very neutron-rich environments, e.g., neutron star mergers. The nucleosynthesis path goes into the very neutron-rich trans-uranium region. The effects of fission are significant to shape the r-process abundances due to fission recycling, of which fission products become seed nuclei ($A < 200$) for the next r-processing during a single nucleosynthesis process. Besides the abundance prediction, fission is also a key role as the heating source of kilonovae at late times ($\sim 10$ days to months) \cite{2018ApJ...868...65W, 2018ApJ...863L..23Z}, electromagnetic transients of neutron star mergers. A sign of fission heating may have been observed in the light curve of the kilonova associated with the gravitational wave, GW170817. The precise understanding of fission becomes much crucial in the era of gravitational astronomy.

In this study, we calculate the fission-fragment mass distributions of very neutron-rich nuclei. Fission product distributions are important for the r-process, but experimental data are not available. We adopt the Langevin method \cite{2019PhRvC.100f4605T, 2019PhRvC..99e1601M}, widely adopted in the study of low-energy fission. We found that the calculated fission mass distributions for uranium, of which the charge ($Z$) distribution is obtained by the UCD (unchanged charge distribution assumption), can well reproduce experimental data in JENDL (${}^{232}{\rm U}$ to ${}^{232}{\rm U}$). We also found that the fission fragment distribution changes from the two peak feature (asymmetric fission) to the one-peak (symmetric fission) with increasing the neutron number.

The present paper is structured as follows. In Section~\ref{sec-method}, we briefly describe the basics of the Langevin calculations and the UCD assumption. The results of the mass and charge distribution are shown in Section~\ref{sec-results}. Section~\ref{sec-conclusion} is dedicated to the conclusion.

\section{Method}\label{sec-method}

\subsection{Dynamical fission calculations by the Langevin method}

We calculate fission fragment distributions (FFD) following fission dynamics by solving the Langevin equations, based on the fluctuation-dissipation model \cite{2019PhRvC..99e1601M, 2019PhRvC.100f4605T}. In the fission calculations, deformation and fission processes are represented by the ``motion'' of trajectories on the multi-dimensional deformation space. For the nuclear shape, we adopt the two-centre-shell model with three parameters (see, \cite{2019PhRvC..99e1601M, 2019PhRvC.100f4605T} for the details), i.e., the distance of two nuclei  ($z_0$), the degree of deformation of a fragment ($\delta$), and the mass asymmetry ($\alpha$).

We calculate the dynamical process of fission by solving the multi-dimensional Langevin equations of motion, which are expressed as follows:
\begin{equation}
\frac{d q_i}{dt} = (m^{-1})_{ji} P_{j}
\end{equation}
\begin{equation}
\label{eq-2}
\frac{d p_i}{dt} = - \frac{\partial V}{\partial q_i} - \frac{1}{2}\frac{\partial}{\partial q_i} (m^{-1})_{jk} P_j P_k - \gamma_{ij} (m^{-1})_{jk} P_k + g_{ij} R_{j}(t) \ ,
\end{equation}
where $q_i = \{z, \sigma, \alpha\}$ and their momentum conjugate, $p_i = m_{ij} dq_i/dt$. $V$ is the potential, and $m_{ij}$ and $\gamma_{ij}$ are the inertia mass and friction coefficient depending on the shape of the nucleus.

The last term of Eq.~\ref{eq-2} corresponds to the random force with the normalized white noise $R_j(t)$. The $g_{ji}$ is the strength of the random force, related with the friction tensor $\gamma_{ji}$ by the classical Einstein relation,
\begin{equation}
\gamma_{ji} T = \sum_k g_{ik} g_{jk}
\end{equation}
where T is the temperature of the nucleus. Following this random property in the momentum equation (Eq.~\ref{eq-2}), each event shows a different trajectory on the potential energy space. We repeat fission dynamics calculations to find the convergence of fission distributions, e.g., the FFD.

\subsection{Charge distributions}
\label{sec-ucd}

Although the time evolution and the FFD are determined in our Langevin calculations, the charge distribution of fission fragments cannot be obtained. However, the charge distribution of fission products is a fundamental quantity for the determination of the production rate of individual isotopes. Thus, it is essential nuclear data to compare experiments with r-process calculations. We calculate the charge distribution based on the assumption of unchanged charge distribution (UCD). The charge distribution (charge density) remains unchanged during the whole fission process, i.e., the charge density of the compound nucleus is maintained.

\begin{figure}[t]
  \centering
  \includegraphics[width=0.8\hsize]{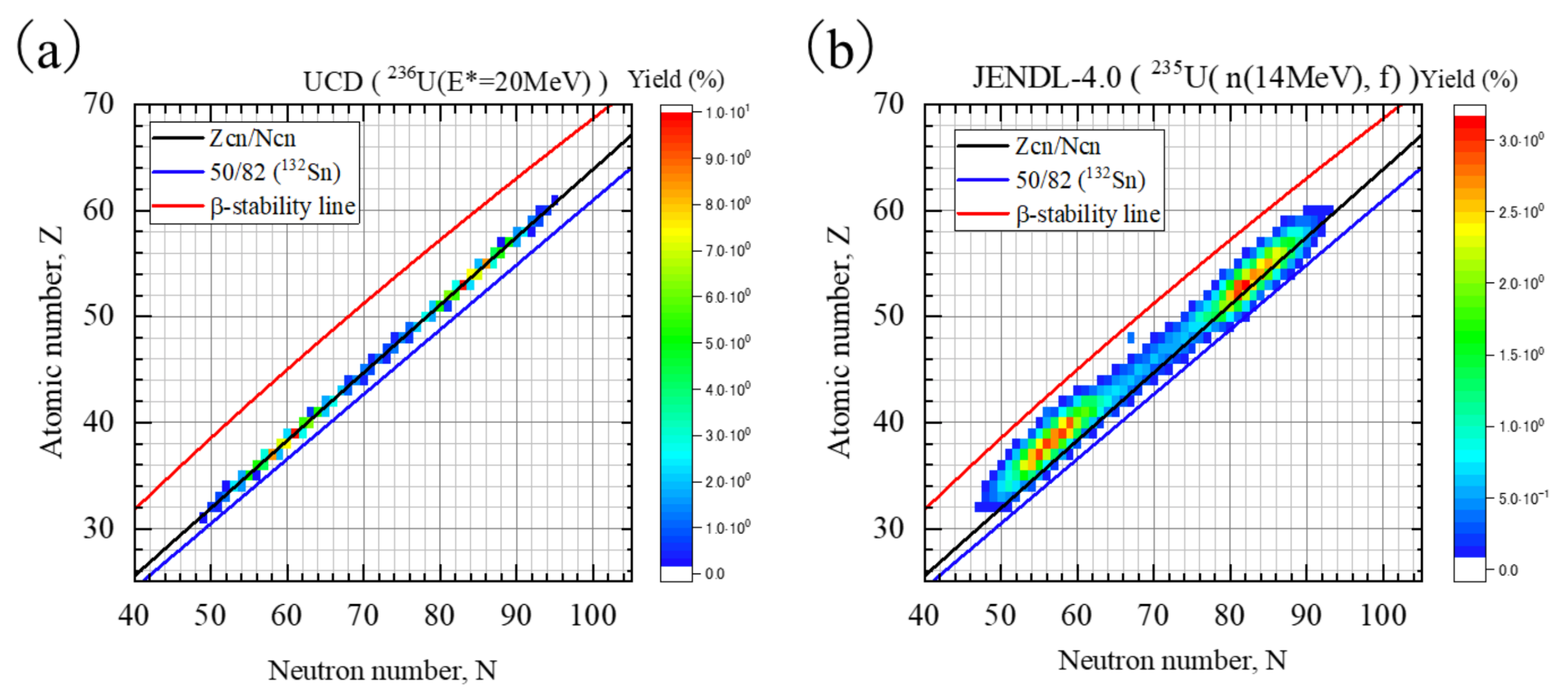}
  \caption{\label{fig-ucd}(a) The calculated FFD on the $N$-$Z$ plane for ${}^{236}{\rm U}$ with the excitation energy: $E^*=20 {\rm MeV}$. (b) The experimental FFD (JENDL-4.0 \cite{2011JKPS...59.1046S}) for ${}^{235}{\rm U}$ by the neutron-induced fission. The $N/Z$ line of compound nucleus (black), the $N/Z = 50/82$ line of a double-magic nucleus ${}^{132}{\rm Sn}$ (blue), and the $\beta$-stability line (red) are plotted.}
\end{figure}

Based on the UCD assumption, we can only calculate the charge ($Z$) distribution along with the $N/Z$ line, i.e., ``projection'' to the $N/Z$ axis. For example, the calculated charge distribution of ${}^{236}{\rm U}$ is shown in Fig.~\ref{fig-ucd}a. However, the distribution must be the width on this line, as shown in the experimental data (Fig.~\ref{fig-ucd}b) of the FFD for neutron-included fission of ${}^{236}{\rm U}$ (the compound nucleus is ${}^{236}{\rm U}$). Therefore, we adopt Gaussian distribution to calculate the width of FFD with $\sigma = 0.01$, relatively well reproduced available experimental data in JENDL-4.0 \cite{2011JKPS...59.1046S}.

\begin{figure}[b]
  \centering
  \includegraphics[width=0.45\hsize]{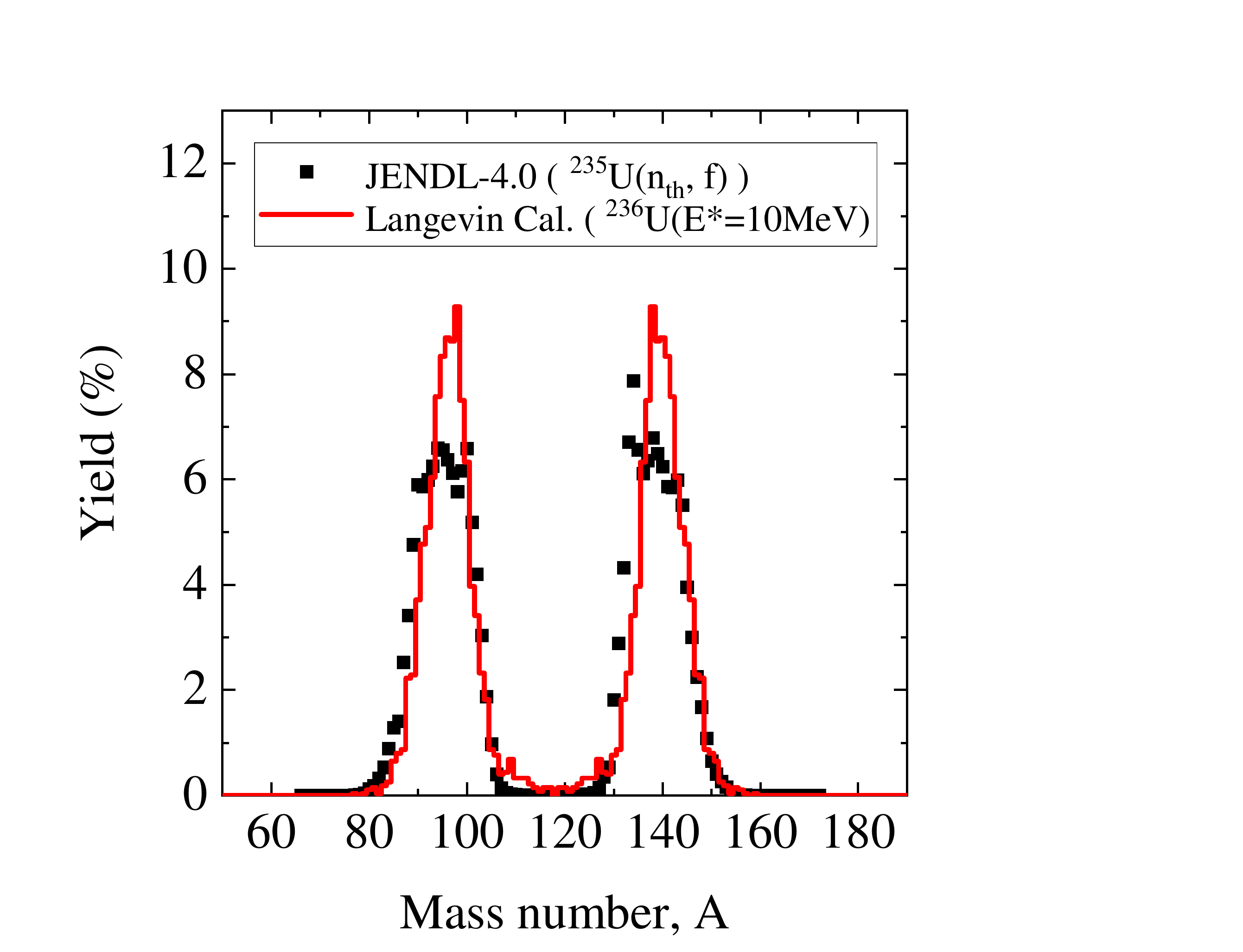}
  \includegraphics[width=0.45\hsize]{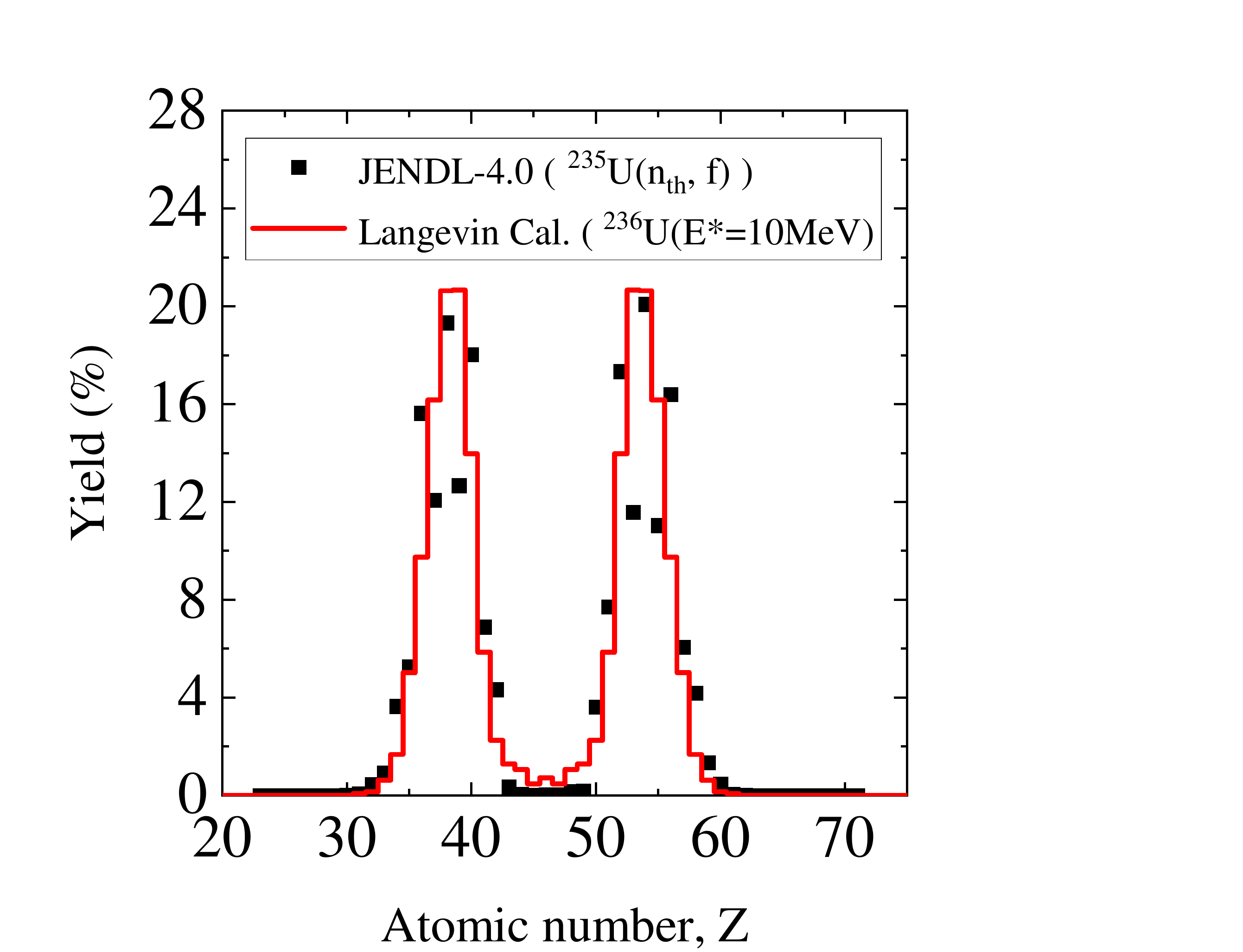}
  \caption{\label{fig-u236}The calculated mass distribution {\it (left)} and the charge distribution {\it (right)} of ${}^{236}{\rm U}$, compared with the experimental data in JENDL-4.0 \cite{2011JKPS...59.1046S}.}
\end{figure}

\section{Results}\label{sec-results}

In this section, we show the fission fragment mass and charge distributions for ${}^{236}{\rm U}$, of which experimental data are available (Section~\ref{sec-u-ffd}). We see our fission model can reproduce experiments well, at least in the less neutron-rich U isotopes. Then, the results of fission calculations for series of U and Fm isotopes are shown in Section~\ref{sec-ffds}.

\subsection{Fission fragment mass and charge distributions}\label{sec-u-ffd}

For the application to the r-process simulations, we need fission properties for very neutron-rich nuclei far from the $\beta$-stability line. However, experimental data in such region are not available yet. Therefore, we firstly carry out fission calculations for U isotopes near the $\beta$-stability line.

Fig.~\ref{fig-u236} shows the results of the mass and charge distribution of fission fragments for ${}^{236}{\rm U}$. The fission calculation is performed with the excitation energy, $E^* = 10~{\rm MeV}$. The distributions are compared with experimental data in JENDL-4.0 \cite{2011JKPS...59.1046S}. The calculated mass distribution, a fundamental variable of the Langevin equations, shows good agreement with experimental data. The charge distribution with the current simple treatment, based on the UCD, also reproduce the experiment.

\begin{figure}[h]
  \centering
  \includegraphics[width=0.45\hsize]{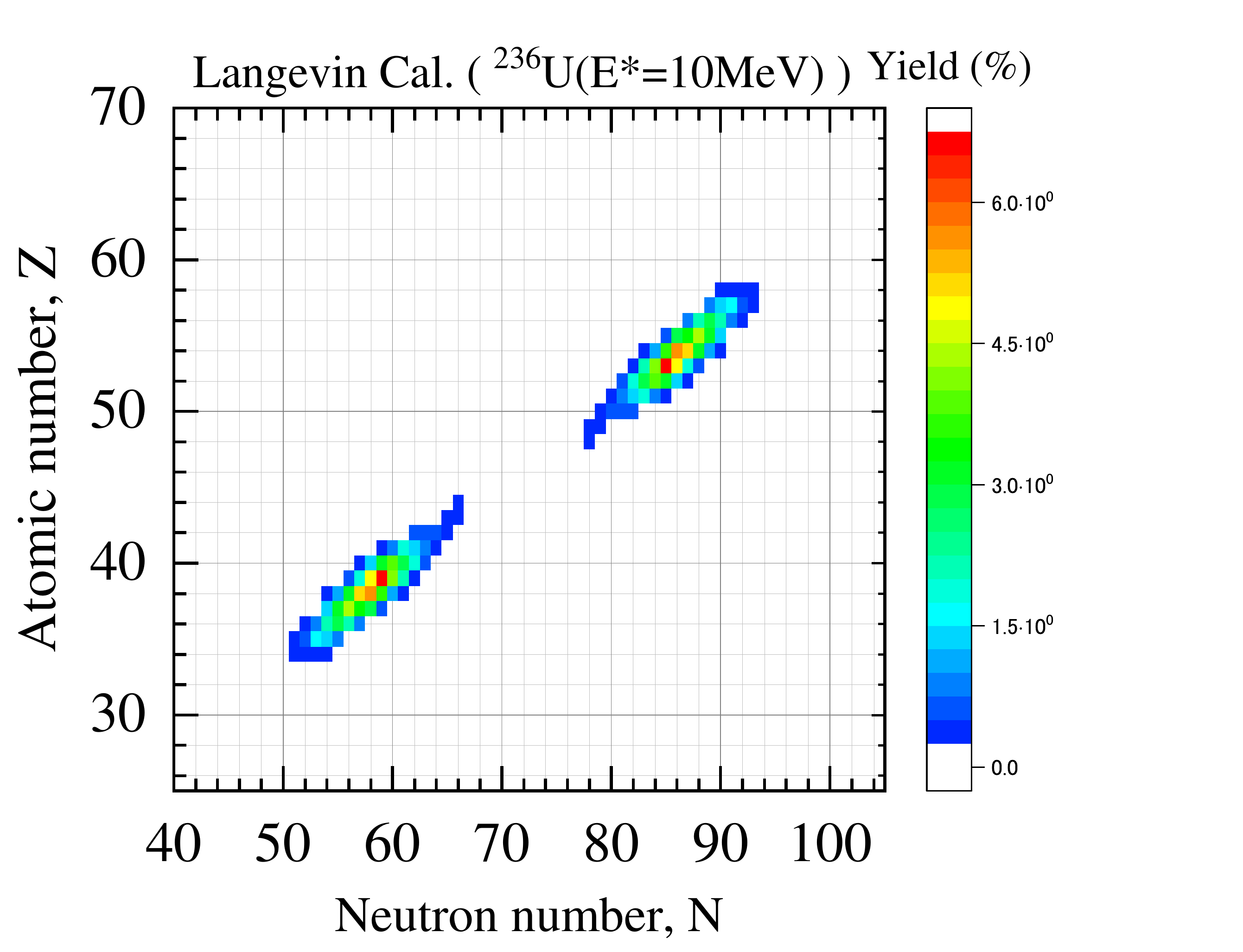}
  \includegraphics[width=0.45\hsize]{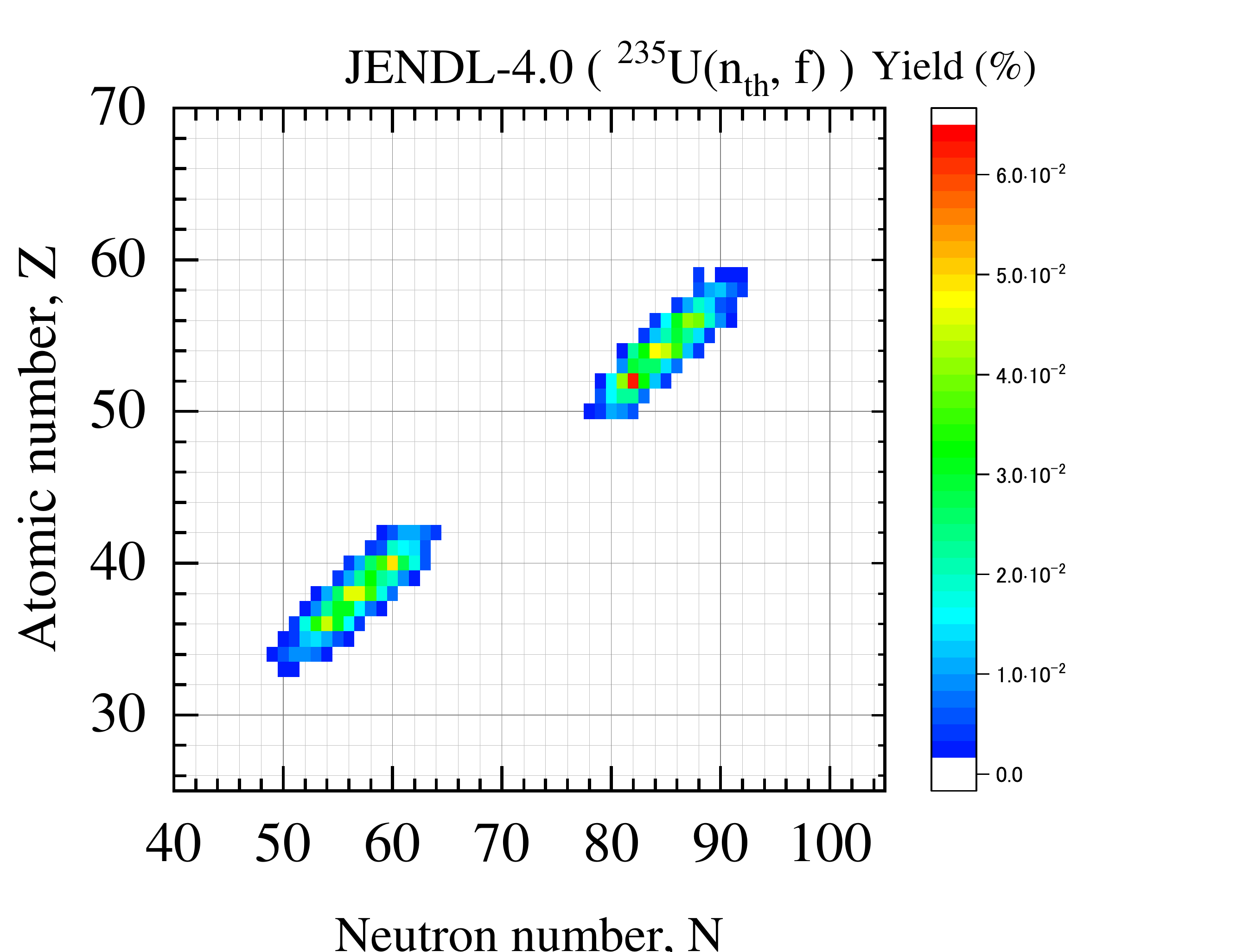}
  \caption{\label{fig-u236-nz}The fission fragment distribution on the $N$--$Z$ plane. The calculation ({\it left}) and experimental data ({\it right}) are plotted.}
\end{figure}

Fig.~\ref{fig-u236-nz} shows the fission fragment distribution on the N-Z plane for ${}^{236}{\rm U}$. The calculation is compared to the neutron-induced fission with the same compound nucleus. The width along the $N/Z$ line is our systematic Gaussian fitting (Section~\ref{sec-ucd}). As expected by Fig.~\ref{fig-u236}, calculated values reproduce basic properties of the JENDL data. 

\begin{figure}[b]
  \centering
  \includegraphics[width=1\hsize]{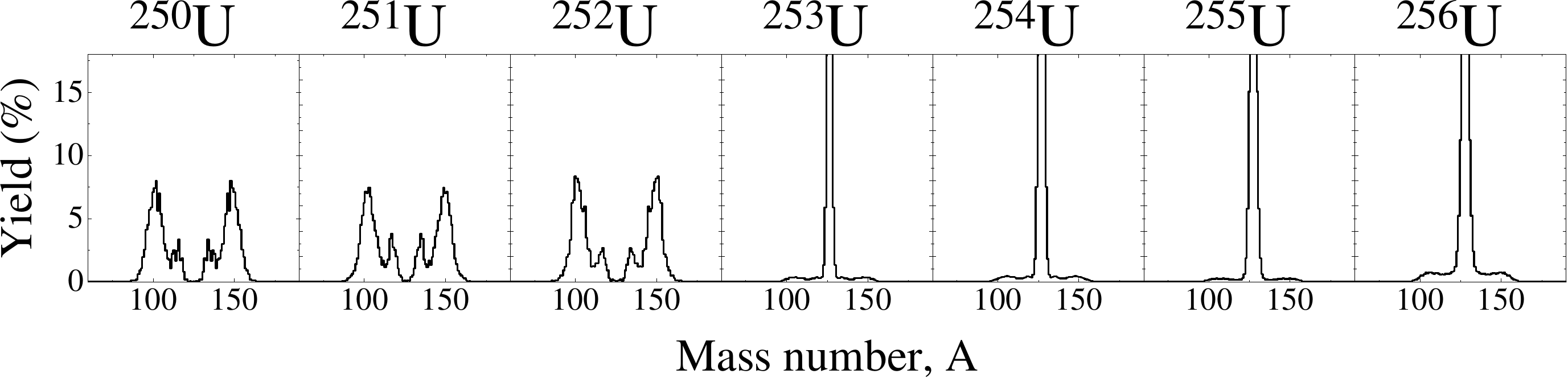}
  \caption{\label{fig-u}The calculated FFD for neutron-rich U ($Z=92$) isotopes, ${}^{250}{\rm U}$ to ${}^{256}{\rm U}$.}
\end{figure}

\subsection{Neutron-rich isotopes for U and Fm}\label{sec-ffds}

We perform a series of fission calculations for the neutron-rich actinoid nuclei. We show the results of U ($Z=92$) and Fm ($Z=100$) isotopes with $10$--$20$ more neutron-rich from the $\beta$-stability line. Fm isotopes are experimentally known that the fission fragmentation becomes asymmetric to symmetric as the neutron number increases. Thus, the number of peaks in FFD changes from two to one in the neutron-rich nuclei.

\begin{figure}[h]
  \centering
  \includegraphics[width=1\hsize]{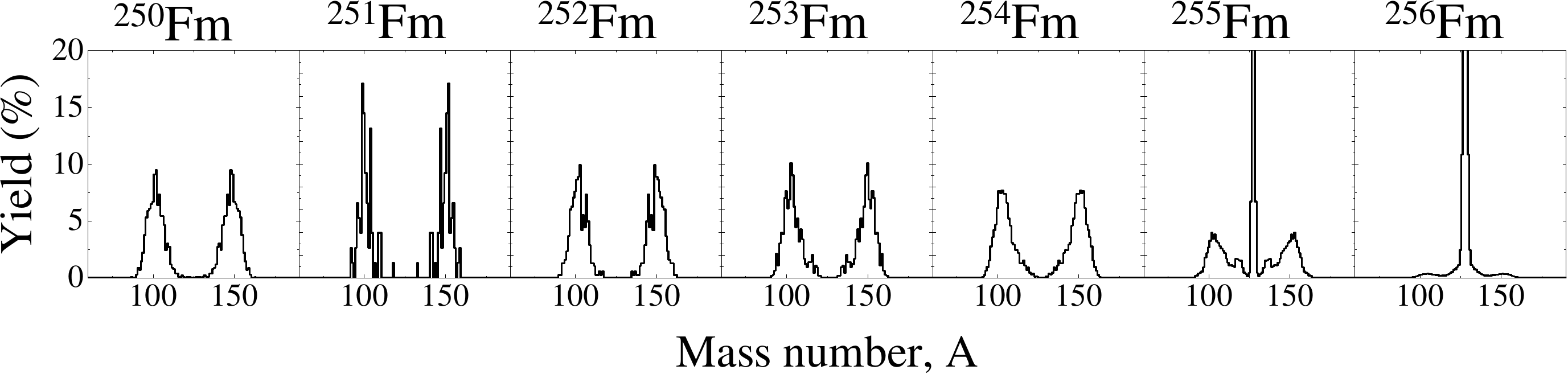}
  \caption{\label{fig-fm}Same as Fig.~\ref{fig-u}, but for Fm ($Z=100$) isotopes.}
\end{figure}

We adopt lower excitation energy, $E^* = 7~{\rm MeV}$, corresponding to low energy environments the r-process occurs. The calculated FFD for ${}^{250}$U to ${}^{256}$U are shown in Fig.~\ref{fig-fm}. We find the distribution changes between ${}^{252}$U and ${}^{253}$U. The number of peaks reduced from two to one, i.e., the FFD becomes symmetric from asymmetric.

We also find the transition of the symmetry in FFD for Fm isotopes. Fig.~\ref{fig-fm} shows mass distributions of ${}^{250-256}{\rm Fm}$. A drastic change in the FFD is found between ${}^{254}{\rm Fm}$ and ${}^{255}{\rm Fm}$. As this set of Fm nuclei is less neutron-rich compared to U, this feature has been suggested by previous experiments.

This transition was found by previous works in the trans-uranium region (see, \cite{2019PhRvC..99e1601M} and references therein), but suggested boundary is $A = 257$ line. This disagreement may cause uncertainties in the Langevin models (model parameters) and the difference of initial conditions. Note that Langevin calculations, based on the same numerical code as the current study, can reproduce the experiments of Fm with specific input physical parameters \cite{2019PhRvC..99e1601M}.

The transition is commonly found in our calculations for U to Db isotopes (Tanaka~et~al., in prep.). Such systematic behaviour can be significant to shape the final abundances of r-process calculations. Conversely, the comparison of r-process calculations with fission distributions can restrict the fission mechanism. In previous studies, such discussions have been done with simplified fission models. Based on dynamical fission calculation, our approach can shed light on the microscopic features of heavy nuclei.

\section{Conclusion}
\label{sec-conclusion}

In the presented work, we have performed fission calculations of neutron-rich nuclei using a kinetic model for application to r-process simulations. We showed the unique property that the mass number distribution changes dramatically with one or two different mass numbers. Such a difference in the distribution affects the production of elements in the synthesis of elements in space. Our results, including further improvements, are expected to contribute to the understanding of the r-process.

As for the Fm region and the neutron-rich nuclides in the U region, experimental data have been obtained by domestic research institutes, including Japan Atomic Energy Agency and RIKEN. It is expected to verify the theoretical calculations. As for the Fm region and the neutron-rich nuclides in the U region, experimental data have been obtained by domestic research institutes, including Japan Atomic Energy Agency and RIKEN RIBF. It is expected to verify the theoretical calculations. To develop a complete theoretical set of fission data, the supporters of such future experiments are desirable.
\\

This study is financially supported by JSPS KAKENHI (19H00693, 20K04003, 20H05648, 21H01087). Parts of computations are performed on computer facilities at CfCA in NAOJ and at YITP in Kyoto University.

\bibliography{ref.bib}

\end{document}